\documentstyle[aps,twocolumn,epsfig]{revtex}
\begin{document}
\newcommand{\ve}[1]{\mbox{\boldmath $#1$}}
\twocolumn[\hsize\textwidth\columnwidth\hsize
\csname@twocolumnfalse%
\endcsname

\draft

\title {Hysteresis and persistent currents in a rotating Bose-Einstein 
condensate with anharmonic confinement}
\author{A. D. Jackson$^1$ and G. M. Kavoulakis$^2$}
\date{February 13, 2004}
\address{$^1$Niels Bohr Institute, Blegdamsvej 17, DK-2100 Copenhagen \O,
             Denmark \\
         $^2$Mathematical Physics, Lund Institute of Technology, P.O.  Box
             118, S-22100 Lund, Sweden}
\maketitle

\begin{abstract}

We examine a Bose-Einstein condensate of atoms that rotates in a
quadratic-plus-quartic potential. It is shown that states of different 
circulation can be metastable.  As a result, we demonstrate that the gas 
can exhibit hysteresis as the angular frequency of rotation of the trap 
is varied.  The simplicity of the picture that emerges for small 
coupling strengths suggests that this system may be attractive for studies 
of phase transitions.

\end{abstract}
\pacs{PACS numbers: 03.75.Hh, 03.75.Kk, 67.40.Vs}

\vskip0.5pc]

\section{Introduction}

The phenomenon of hysteresis is wide-spread in physics and can be 
expected whenever a physical system undergoes a first-order phase transition.
Consider, for example, a system whose free energy has two minima as a 
function of its order parameter.  As some external parameter is varied, 
these can shift with respect to each other.  Provided only that fluctuations 
are not of a strength sufficient to bring the system to the absolute 
minimum of its free energy, it will remain at a local minimum even of this 
is not the absolute minimum, i.e., even if it is in a metastable 
state. The existence of metastable states implies that the actual 
state of the system depends on its history.

Superfluids number among the many physical systems which can display 
hysteresis.  The rigidity against weak perturbations of the state of 
a superfluid that rotates in a bucket or flows in a pipe can be attributed to 
its metastability \cite{Leggett}.  For this reason, these fluids are
expected to show hysteresis and actually do so.  For example, if such
a fluid is inside a container which slowly starts to rotate, it remains
stationary until the angular frequency reaches some critical value. In the 
reverse process, with the fluid in motion, the system follows a different
path, and the fluid continues to rotate even when the container does not.

In the present paper we demonstrate how these ideas apply to a Bose-Einstein
condensate of alkali-metal atoms that rotates in a quadratic-plus-quartic 
potential. Our motivation for this study comes from recent experiments 
in such trapping potentials which observed vortex states \cite{Dal,Dal2}. 
Using a simple but still realistic model \cite{JK,JKL}, we demonstrate 
in Sec.\,II the existence of metastable states in these systems. In Sec.\,III
we examine how the effects of hysteresis can appear as the frequency of 
rotation of the trap changes. In Sec.\,IV we discuss our results, and 
Sec.\,V summarizes the basic points of our study.

\section{Model -- Structure of the energy surface}

To see how hysteresis works in this problem, it is useful to give
a brief summary of the results of Refs.\,\cite{JK,JKL}. One of the most
important elements of the present study is to understand the structure of the
energy of the gas in the rotating frame of reference, $E' = E - {\bf L}
\cdot {\bf \Omega}$, where $E$ is the energy in the lab frame,
${\bf L}$ is its angular momentum, and ${\bf \Omega}$ is the frequency of
rotation of the external trapping potential $V(\rho)$.  This trapping 
potential is assumed to have the form
\begin{equation}
   V(\rho) = \frac 1 2 M \omega^2 \rho^2 [1 + \lambda (\frac {\rho} {a_0})^2].
\label{anh}
\end{equation}
Here, $\rho$ is the cylindrical polar coordinate, $M$ is the atomic mass,
$a_0 = (\hbar/M \omega)^{1/2}$ is the oscillator length, and $\lambda$
is a small dimensionless constant.  (In the experiment of Ref.\,\cite{Dal},
$\lambda \approx 10^{-3}$.) The trapping along the axis of rotation, the 
$z$ axis, has been neglected for simplicity, and the appropriate density
is thus the number of atoms $N$ per unit length $Z$, $\sigma = N/Z$. 

The (repulsive) interaction between atoms is assumed to have the usual form 
\begin{eqnarray}
   V_{\rm int} =
      \frac 1 2 U_{0} \sum_{i \neq j} \delta({\bf r}_{i} - {\bf r}_{j}),
\label{v}
\end{eqnarray}
where $U_0 = 4 \pi \hbar^2 a/M$ is the strength of the effective
two-body interaction with $a$ equal to the scattering length for
atom-atom collisions. In the limit of weak interactions considered
here, the typical atomic density $n$ is $\sim N/(\pi a_0^2 Z)$. 
Thus, for a typical interaction energy $n U_0$, $n U_0/\hbar \omega 
\sim \sigma a$ and the dimensionless quantity which plays the role 
of a coupling constant in this two-dimensional problem is $\sigma a$,  
which is assumed to be smaller than unity.

In the limit $\sigma a \ll 1$ and $\lambda \ll 1$ the appropriate basis 
states with angular momentum $m \hbar$ are the eigenstates of the harmonic 
potential with no radial nodes,
\begin{equation}
    \Phi_m({\rho, \phi}) = \frac 1 {(m! \pi a_0^2 Z)^{1/2}}
           \left( \frac {\rho}{a_0} \right)^{|m|} e^{i m \phi}
	           e^{-\rho^{2}/2 a_0^2},
\label{phim}
\end{equation}
where $\phi$ is the angle in cylindrical polar coordinates. Quite generally, 
the order parameter, $\Psi$, of the condensate can then be expanded in this 
basis as 
\begin{equation}
   \Psi = \sum_m c_m \Phi_m.
\label{opr}
\end{equation}
It was shown in Ref.\,\cite{JK} that, for sufficiently small $\sigma a$, 
the energy of the gas in the rotating frame, $E'$, is minimized when only 
one component $m = m_0$ in the expansion of Eq.\,(\ref{opr}) contributes to 
the order parameter, $\Psi = \Phi_{m_0}$. In addition, the critical 
frequencies which denote the lower limit on the {\it absolute\/} stability 
of the states $\Phi_{m_0}$ are given by
\begin{eqnarray}
   \frac {\Omega_{m_0}} {\omega} =
      \frac {m_0} {|m_0|} \left[ 1 + \lambda (|m_0|+1)
      \right. \phantom{XXXXX} \nonumber \\ \left.
     - \sigma a \frac {(2 |m_0| - 2)!} {2^{2 |m_0| - 1} (|m_0|-1)! |m_0|!}
            \right].
\label{gen}
\end{eqnarray}
The solid (straight) line in Fig.\,1 shows this boundary between the states
with $m_0 = 0$ and $m_0 = 1$. On the left of this line the energy $E'$ has
an absolute minimum for $\Psi = \Phi_0$; on the right the absolute 
minimum occurs for $\Psi = \Phi_1$. This is shown schematically 
in Fig.\,2, where the minimum on the left in the graphs corresponds to the 
state $\Phi_0$, while the one on the right corresponds to the state $\Phi_1$.

As discussed in Refs.\,\cite{JK,JKL}, there are metastable states for 
certain values of $\sigma a$ and $\Omega/\omega$ in addition to these 
absolute minima.  Consider, for example, the case $m_0 = 0$ of a non-rotating 
condensate.  (The same procedure can be applied to any other $m_0$.)  While 
the state $\Phi_0$ represents the absolute minimum of $E'$ in regions I and 
II, it is metastable in region III and unstable in region IV with respect 
to the state
\begin{eqnarray}
   \Psi = c_{-1} \Phi_{-1} + c_{0} \Phi_{0} + c_{1} \Phi_{1}
\label{psis}
\end{eqnarray}
for $\sigma a < 0.06$. For $\sigma a > 0.06$ (but $\sigma a$ still 
sufficiently small) the instability is towards the state 
\begin{eqnarray}
   \Psi = c_{-2} \Phi_{-2} + c_{0} \Phi_{0} + c_{2} \Phi_{2}.
\label{psis2}
\end{eqnarray}
These genuine instabilities are given by the right dashed line in Fig.\,1.

To be specific, the energy $E'$ of the system per particle in the 
rotating frame in the state of Eq.\,(\ref{psis}) is
\begin{eqnarray}
  \frac {E'} {N \hbar \omega} = |c_{-1}|^2 (1 + \Omega/\omega)
   + |c_{1}|^2 (1 - \Omega/\omega)
\nonumber \\
      + \lambda (3 |c_{-1}|^2 + |c_0|^2 + 3 |c_1|^2)
\nonumber \\
            + \sigma a (\frac 1 2 |c_{-1}|^4 + |c_{0}|^4 
	    + \frac 1 2 |c_{1}|^4 + 2 |c_{-1}|^2 |c_0|^2  
\nonumber \\
	    + 2 |c_0|^2 |c_{1}|^2 + 2 |c_{-1}|^2 |c_{1}|^2 
- 2 |c_{-1}| |c_0|^2 |c_1|).
\label{psis3}
\end{eqnarray}
Making use of the normalization condition, $E'$ can be expressed in terms 
of $c_0$ and $c_1$, and is a two-dimensional energy surface. As one crosses 
the line $\Omega/\omega = 1 + 2 \lambda - \sigma a/2$ (i.e., the straight 
solid line in Fig.\,1 given by Eq.\,(\ref{gen}) for $m_0 = 0$), the absolute 
minimum moves discontinuously from $|c_0|^2 = 1$ to $|c_1|^2 = 1$. Furthermore,
even above this line, there is a region in the $\Omega/\omega$ -- $\sigma a$
plane (i.e., region III in Fig.\,1), for which $E'$ has a local minimum 
at $|c_0|^2 = 1$.  The state $\Phi_0$ is thus an absolute minimum of  
$E'$ in regions I and II, and it describes a local minimum in $E'$ in 
region III, i.e., it is metastable. In region IV, on the other hand, the 
local minimum disappears. These observations are shown schematically in 
Fig.\,2, where the minimum on the left corresponds to $\Phi_0$.

Similarly, for $\Psi=\Phi_1$, the left dashed line in Fig.\,1 gives the 
boundary for the (local) stability of the state $\Phi_1$ with respect to 
the state
\begin{eqnarray}
   \Psi = c_{0} \Phi_{0} + c_{1} \Phi_{1} + c_{2} \Phi_{2}.
\label{psit}
\end{eqnarray}
In regions III and IV, $\Phi_1$ represents the absolute minimum in $E'$, 
in region II it is metastable, and in region I it no longer represents 
a local minimum. These facts are also shown schematically in Fig.\,2, where
the minimum on the right corresponds to $\Phi_1$.

The general picture of stability/metastability that emerges for all 
$m_0 \ge 2$ is shown as the dashed line in Fig.\,3 for $m_0 = 2$. The state
$\Phi_2$ provides the absolute minimum of $E'$ between the straight lines, 
while in the remaining part, $\Phi_2$ is metastable. Along the dashed curve, 
and as one moves clockwise, the instability is against a linear combination 
of states involving the following $m$: $(1,2,3), (0,2,4), (-2,2,6), (-1,2,5), 
(0,2,4)$, and $(1,2,3)$.

\section{Hysteresis}

With these observations, it is easy to demonstrate the effect of hysteresis.
Assume that the condensate is initially at rest and that $\Omega$ increases 
slowly from zero with $\sigma a$ kept fixed (and less than 0.06 for 
simplicity.)  So long as $\Omega/\omega$ is in the regions I and II, 
the cloud will not rotate since $E'$ has its absolute minimum in the state 
with $m_0 = 0$. Provided that fluctuations are sufficiently weak, the 
system will not rotate even in region III since $\Phi_0$ is a metastable 
state. The gas will begin to rotate only when $\Omega/\omega$ is in region 
IV, where $\Phi_0$ is unstable against the state of Eq.\,(\ref{psis}).  The 
system will then move to the pure state $\Phi_1$ (i.e., the absolute minimum 
of the system) on the fastest available time scale.  In this process, it 
is important to note that the gas is unstable with respect to the state
of Eq.\,(\ref{psis}) in region IV.  Since $\Phi_1$ is one of the components 
of $\Psi$, there is a monotonically-decreasing path in the energy from the 
(unstable) pure state $\Phi_0$ to the pure state $\Phi_1$, which is the 
absolute minimum of $E'$.

If one now considers the inverse process in which $\Omega$ decreases slowly 
from region IV with the gas initially in the state $\Phi_1$, the system 
will remain in this state in regions IV and III where it is the absolute 
minimum and in region II where $\Phi_1$ is metastable.  In the absence of
\noindent
\begin{figure}
\begin{center}
\epsfig{file=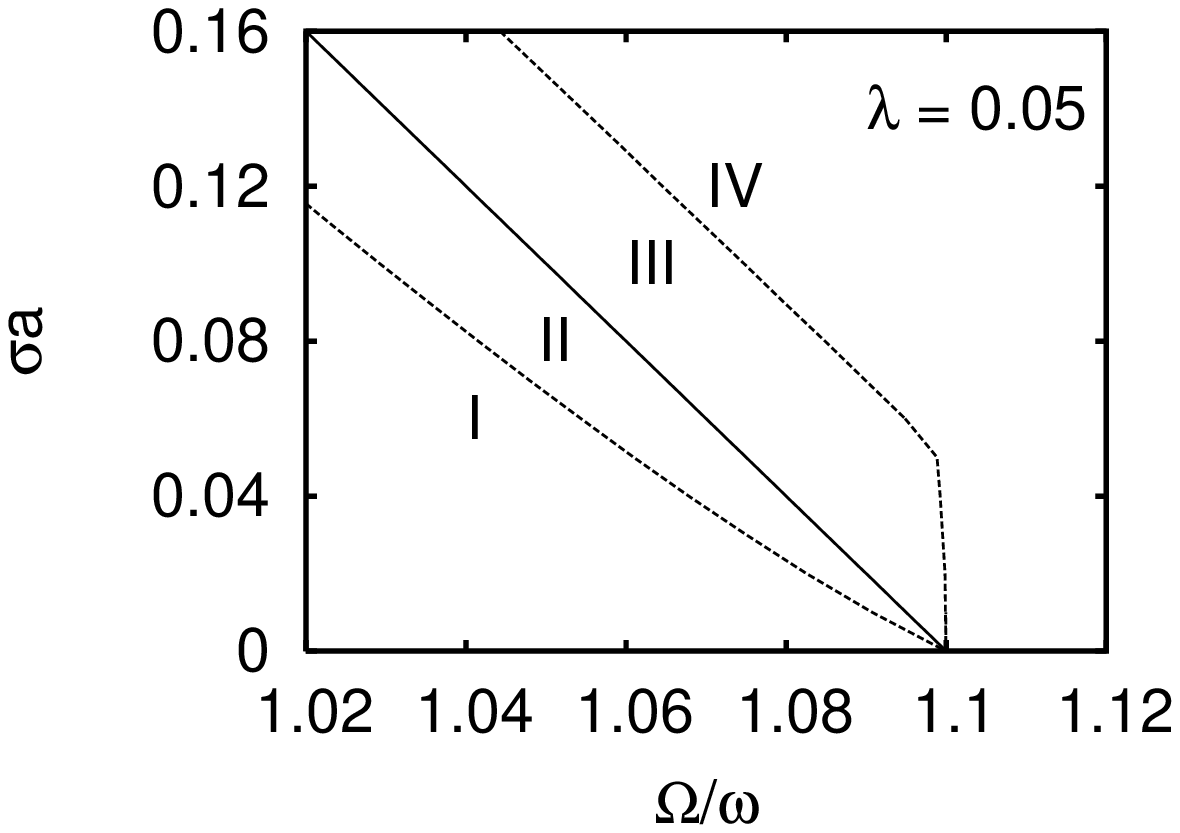,width=8.5cm,height=6.2cm,angle=0}
\vskip0.5pc
\begin{caption}
{The phase diagram of a rotating Bose-Einstein condensate
confined in a quadratic-plus-quartic potential in the $\Omega/\omega$ --
$\sigma a$ plane. The straight solid line gives the phase boundary between
the phases where the states $\Phi_0$ (regions I, II) and $\Phi_1$
(regions III, IV) provide the absolute minima of the energy in the rotating
frame. $\Phi_0$ is metastable in region III and unstable in region IV.
Similarly $\Phi_1$ is metastable in region II and unstable in region I.}
\end{caption}
\end{center}
\label{FIG1}
\end{figure}
\noindent
significant fluctuations, a transition will occur only when
$\Omega/\omega$ reaches region I.  In this region, $\Phi_1$ is unstable to the 
state of Eq.\,(\ref{psit}), which includes a component of $\Phi_0$.  The system 
will move to the absolute minimum described by the (pure) state $\Phi_0$, 
and rotation will cease.  The gas thus displays hysteresis in regions II 
and III.  The order parameter is given by $\Phi_0$ as $\Omega$ increases; 
by $\Phi_1$ as it decreases.

A similar process evidently exists for larger values of $\Omega$, 
and we show the first few steps of the angular momentum $m$ of the gas as a 
function of $\Omega/\omega$ in Fig.\,4 for $\lambda = 0.05$ and $\sigma a = 
0.05$. The picture that emerges for small values of $\sigma a$ is clear, since 
the leading instability of a given state $\Phi_{m_0}$ (for 
increasing/decreasing $\Omega$) always involves a linear superposition of 
the form 
\begin{eqnarray}
   \Psi = c_{m_0 - n} \Phi_{m_0 - n} + c_{m_0} \Phi_{m_0} 
     + c_{m_0 + n} \Phi_{m_0 + n},
\label{gfor1}
\end{eqnarray}
with $n = 1$. (For $\sigma a = 0$ the equation that yields each solid curve 
coincides with those leading to the corresponding dashed curves \cite{JK,JKL}.  
Thus, the two dashed curves always meet their solid partner at $\sigma a = 
0$.) As soon as $\Phi_{m_0}$ becomes unstable, the order parameter will 
adjust rapidly to the new minimum provided by the state $\Phi_{m_0 \pm 1}$. 

We have found that the largest value of $\sigma a$, $(\sigma a)_c$, for which 
the instability of a given $\Phi_{m_0}$ is to a state of the form of 
Eq.\,(\ref{gfor1}) with $n=1$ increases roughly linearly with $m$ [$(\sigma a)_c 
= 0.06$ for $m_0 = 0$]. In this case $E'$ can have at most two local minima.  The
transition between different states as $\Omega$ increases/decreases is then clear 
and consists of steps in the angular momentum $m$ of the gas equal to $\pm 1$.
For fixed $\sigma a$, the pattern shown in Fig.\,4 
\noindent
\begin{figure}
\begin{center}
\epsfig{file=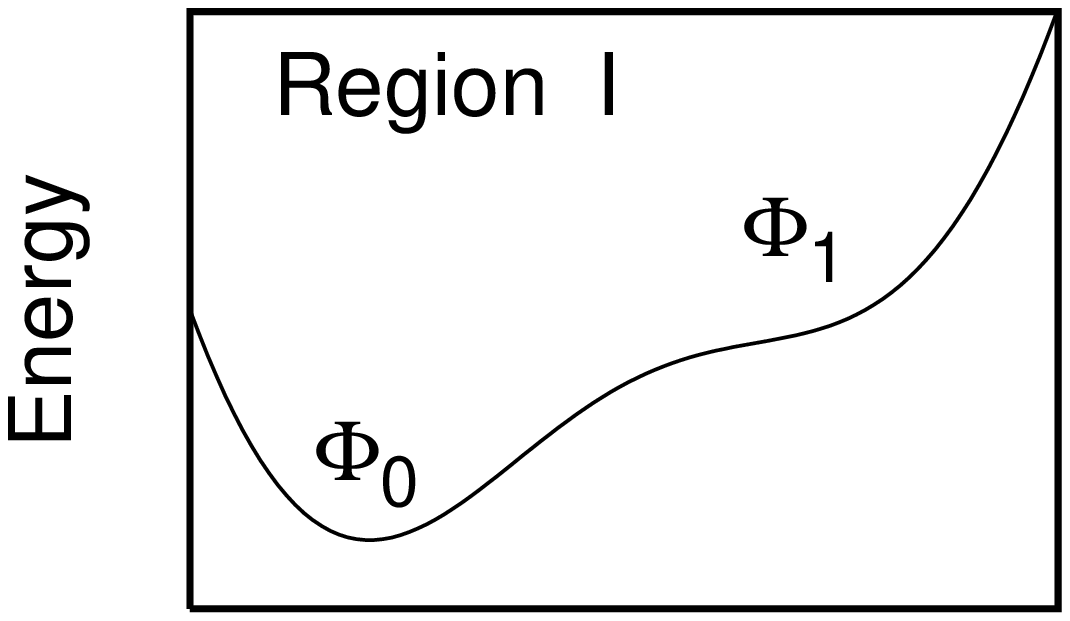,width=3.5cm,height=3.1cm,angle=0}
\epsfig{file=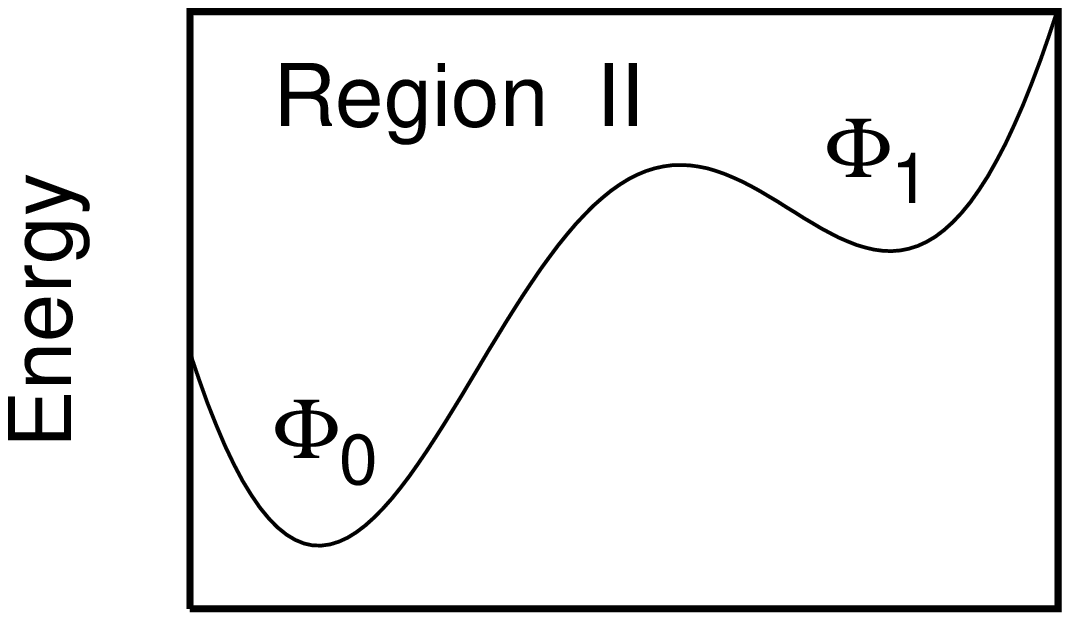,width=3.5cm,height=3.1cm,angle=0}
\epsfig{file=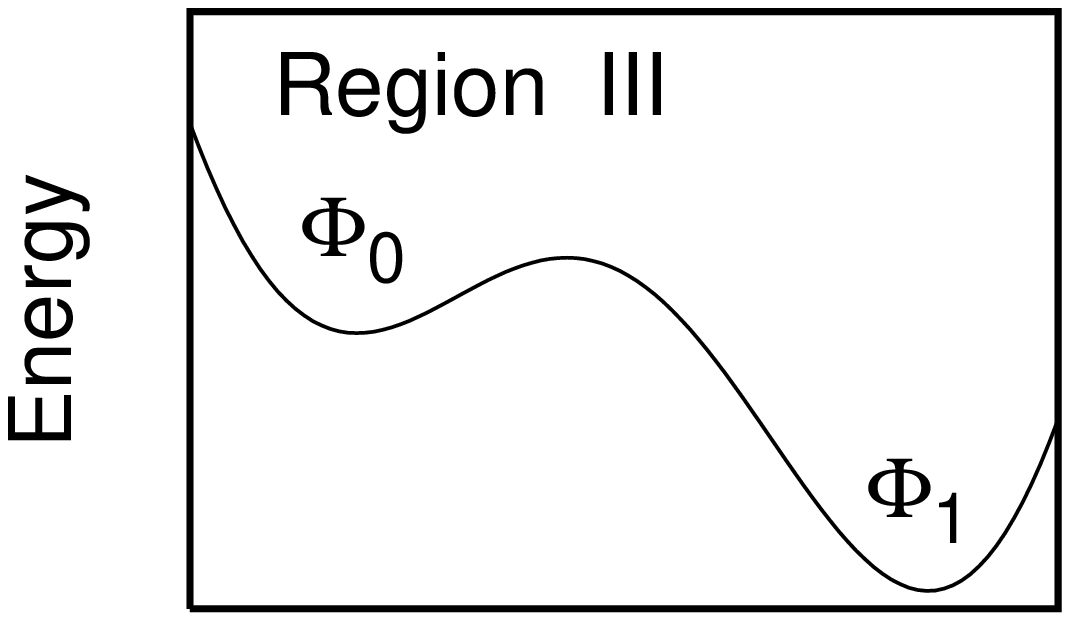,width=3.5cm,height=3.1cm,angle=0}
\epsfig{file=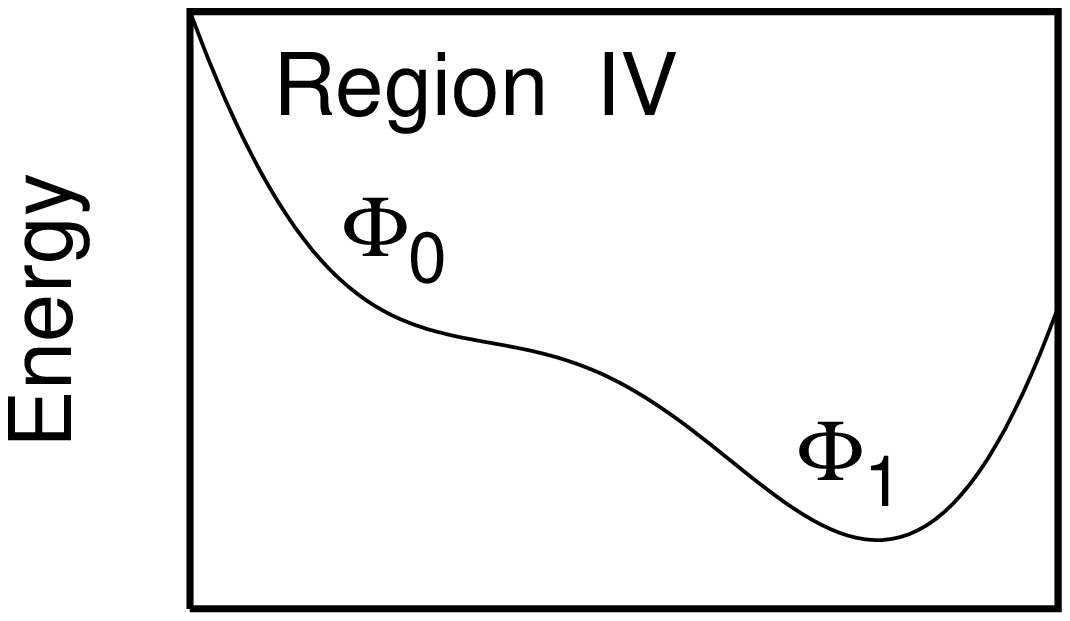,width=3.5cm,height=3.1cm,angle=0}
\vskip0.5pc
\begin{caption}
{Schematic diagram of the energy of the system in the rotating frame as
a function of the order parameter for a fixed $\sigma a$ and for values of
$\Omega/\omega$ in regions corresponding to those in Fig.\,1. The minimum
on the left corresponds to the state $\Phi_0$; that on the right to
the state $\Phi_1$.}
\end{caption}
\end{center}
\label{FIG2}
\end{figure}
\noindent
will always hold if $\Omega/\omega$ is sufficiently large since $(\sigma a)_c$ 
increases with $m$ and $\sigma a$ will eventually become smaller than $(\sigma a)_c$.

For values of $\sigma a > (\sigma a)_c$ the situation becomes more rich,
since there are instabilities to many different combinations of states, 
depending on the specific value of $\sigma a$ [like, for example, the one in
Eq.\,(\ref{psis2})] of the general form given in Eq.\,(\ref{gfor1}),
with $n$ being a integer. In this case there is a wide variety of possibilities
where the system will evolve as $\Omega$ changes. The (multi-dimensional)
energy $E'$ develops many local minima and small changes in $\sigma a$ may 
drive the gas to completely different states, as it will follow the path 
with the steepest decrease in its energy. The corresponding graph of Fig.\,4
is then in principle more complicated, as the steps in $|m|$ may be different
than unity and can be non-integer.  When the absolute minimum energy is 
given by Eq.\,(\ref{gfor1}) with all coefficients non-zero, the order 
parameter has an $n$-fold {\it discrete\/} rotational symmetry \cite{JKL}.  
This observation is of some importance.  The existence of metastable states 
is a necessary condition for hysteresis.  As repeatedly noted, the ability 
to observe hysteresis also requires the absence of fluctuations sufficient 
to cause phase transitions from metastable states.  The fluctuations required 
to change discrete rotational symmetry become relatively improbable and 
the experimental indications of hysteresis are likely to be clearer for 
larger values of $\sigma a$ where one or both states are given by 
Eq.\,(\ref{gfor1}) \cite{JKL}.

\section{Discussion}

Reference \cite{PG} has shown that similar effects of hysteresis also 
appear in harmonically-trapped Bose-Einstein 
\noindent
\begin{figure}
\begin{center}
\epsfig{file=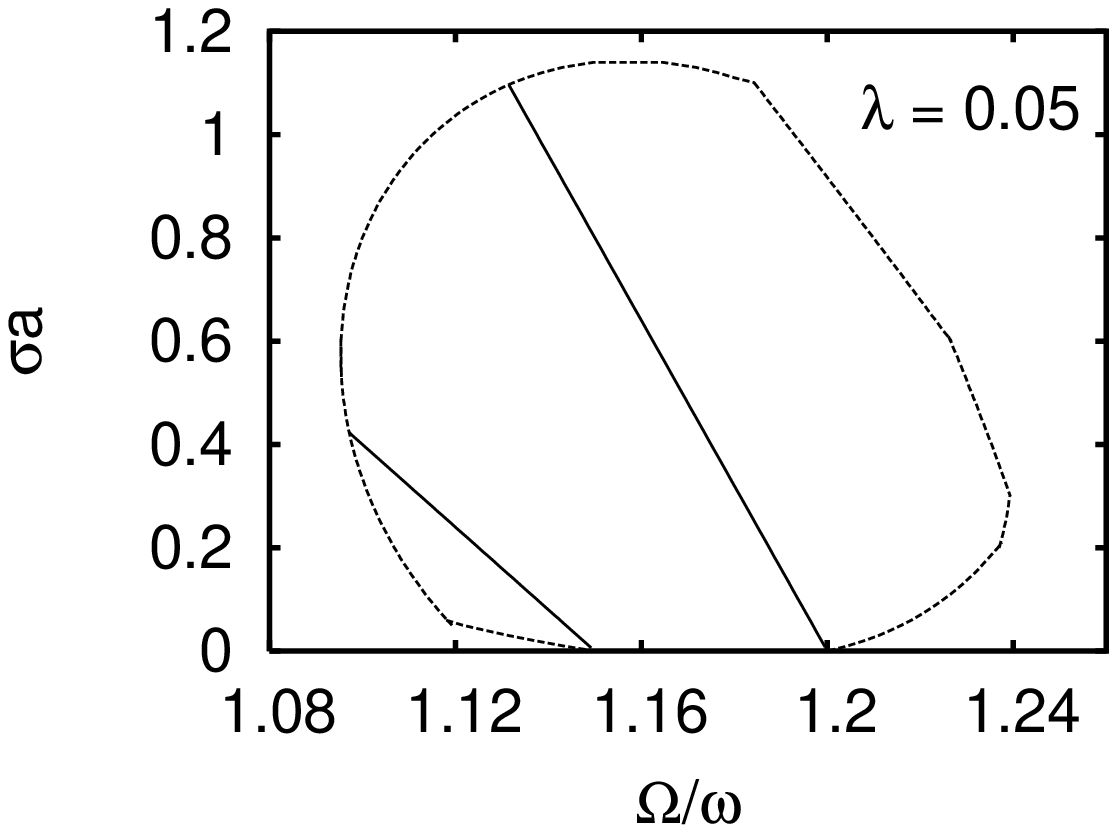,width=8.0cm,height=6.0cm,angle=0}
\vskip0.5pc
\begin{caption}
{Regions of absolute stability/metastability of the state $\Phi_2$. Between 
the solid lines $\Phi_2$ provides the absolute minimum of $E'$, while in the 
remaining region enclosed by the dashed curve it is metastable.}
\end{caption}
\end{center}
\label{FIG3}
\end{figure}
\noindent
condensates.  (See also \cite{EM}.)  This is no surprise since hysteresis is 
a general feature of superfluids. These results are also consistent 
with the corresponding experimental observations of Ref.\,\cite{Madison}.

One important aspect of our study, which also applies in the case of harmonic
trapping, is that the critical frequencies for the creation of vortex states 
observed in such experiments depend on how the experiment is performed.  For 
example, if one first cools the gas in the condensed phase and then rotates 
the trap, the picture that emerges is that of Fig.\,4.  If, on the other 
hand, the trap is first set in motion and the gas then cooled, it will
end up at the absolute minima of $E'$ as given by Eq.\,(\ref{gen}).  (See 
Fig.\,1 in Ref.\,\cite{JK}.)

The simple structure of $E'$ for small values of $\sigma a$ described 
here is an interesting feature, and this system could prove to be an 
ideal laboratory for the investigation of the effects of quantum and thermal 
fluctuations on phase transitions. It is important to mention that the 
boundaries we have calculated here are exact in the limit of small 
$\sigma a$. As in other problems that involve phase transitions between 
metastable states, it is assumed that fluctuations are suppressed because of 
the energy barrier that separates the two states.  In the presence of 
such an energy barrier, the characteristic time scale for the phase 
transition is expected to be long --- particularly when the two local 
minima have distinct discrete symmetry.  When the barrier disappears and the 
metastable state becomes unstable, the phase transition proceeds rapidly.  
Here, we have suggested an essentially exact description of the phase 
boundaries and the corrersponding energy surface.

The effects considered here could be useful for (i) probing persistent
currents in superfluids, (ii) providing experimental information regarding 
the energy surface of the system, and (iii) investigating the effect of 
fluctuations on the phase transitions (e.g., by varying either the number 
of atoms or the temperature).
\noindent
\begin{figure}
\begin{center}
\epsfig{file=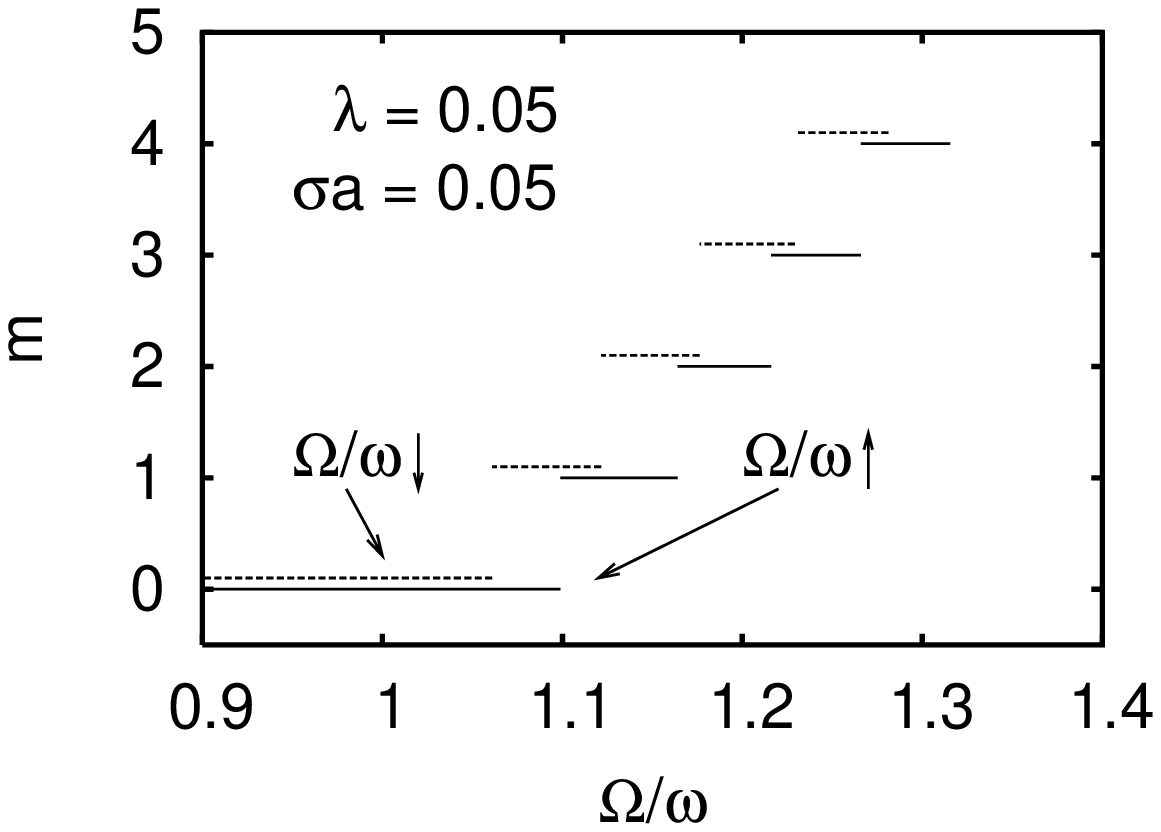,width=8.0cm,height=6.0cm,angle=0}
\vskip0.5pc
\begin{caption}
{Demonstration of hysteresis. The graph shows the angular momentum, $m$,
of the system as function of the frequency of the rotation of the trap,
$\Omega/\omega$, as $\Omega/\omega$ increases (solid lines) and as it
decreases (dashed lines).  The dashed lines have been displaced slightly.
Here, $\lambda = 0.05$ and $\sigma a = 0.05$.}
\end{caption}
\end{center}
\label{FIG4}
\end{figure}
\noindent
\section{Summary}

In summary, we have investigated the energy of a
Bose-Einstein condensate that rotates in a quadratic-plus-quartic trapping
potential. The energy surface is elementary in the limit of small
coupling with a pattern of absolutely stable and metastable states which 
can give rise to potentially observable persistent currents and hysteresis 
phenomena. Finally, because of its simplicity, the present system can 
offer some insight into the physics of phase transitions that are driven 
by fluctuations.

\end{document}